# Controlling Curie temperature in (Ga,Mn)As through location of the Fermi level within the impurity band

M. Dobrowolska[1,*], K. Tivakornsasithorn[1], X. Liu[1], J. K. Furdyna[1], M. Berciu[2], K. M. Yu[3] and W. Walukiewicz[3]

[1]Department of Physics, University of Notre Dame, Notre Dame, IN  46556, USA

[2]Department of Physics and Astronomy, University of British Columbia, Vancouver, BC V6T 1Z1, Canada

[3] Materials Science Division, Lawrence Berkeley National Laboratory, Berkeley, California, 94720, USA

## Abstract

The ferromagnetic semiconductor (Ga,Mn)As has emerged as the most studied material for prototype applications in semiconductor spintronics. Because ferromagnetism in (Ga,Mn)As is hole-mediated, the nature of the hole states has direct and crucial bearing on its Curie temperature $T_C$. It is vigorously debated, however, whether holes in (Ga,Mn)As reside in the valence band or in an impurity band.  In this paper we combine results of channeling experiments, which measure the concentrations both of Mn ions and of holes relevant to the ferromagnetic order, with magnetization, transport, and magneto-optical data to address this issue. Taken together, these measurements provide strong evidence that it is the location of the Fermi level within the impurity band that determines $T_C$ through determining the degree of hole localization. This finding differs drastically from the often accepted view that $T_C$ is controlled by valence band holes, thus opening new avenues for achieving higher values of $T_C$.

*Email: mdobrowo@nd.edu

Understanding the factors that control the Curie temperature $T_C$ in (Ga,Mn)As is of obvious importance, since it can serve as a guide for strategies to optimize this material. This issue is closely linked to the question of whether the holes mediating the Mn-Mn interaction reside in a weakly disordered valence band, or in an impurity band. Despite extensive studies, this question is still vigorously debated[1]. The valence band model assumes that a separated impurity band does not exist for Mn concentrations higher than ~1 to 2%[2,3], and successfully accounts for a number of observations[2-12]. The alternative model assumes that the holes reside in a Mn-derived impurity band even for moderate to high Mn concentrations[13-15], and there is also a wealth of experimental papers favoring this picture[16-26].

When Mn ions substitute for Ga in (Ga,Mn)As, they become acceptors, introducing holes that mediate ferromagnetic interactions between the $S = 5/2$ moments of the half-filled $3d$ Mn shells. However, as was described theoretically[16] and observed experimentally[27], during (Ga,Mn)As growth some of the Mn ions also enter into interstitial sites ($Mn_I$), becoming highly-mobile positively charged double donors. They can, however, be immobilized by Coulomb attraction at interstitial sites immediately adjacent to the negatively-charged $Mn_{Ga}$, where they form antiferromagneticaly-coupled $Mn_I$-$Mn_{Ga}$ pairs, as was shown both theoretically[28,29] and experimentally[30-32]. This automatically lowers the fraction of Mn contributing to ferromagnetic order to $x_{eff} = x_{sub} - x_I$, where $x_{sub}$ and $x_I$ are respectively the concentrations of $Mn_{Ga}$ and $Mn_I$. Since $Mn_I$ are double donors, they also reduce the concentration of the holes by compensation to $p = 4(x_{sub} - 2x_I)/a^3$, where $a$ is the lattice constant of (Ga,Mn)As.

The early mean field theory based on the assumption that holes reside in the valence band predicts[2] that $T_C \sim x_{eff} p^{1/3}$. Later this theory was expanded by Jungwirth *et al.*[33] in order to include additional refinements, such as discreteness of random $Mn_{Ga}$ positions in the lattice and antiferromagnetic superexchange contributions to the near-neighbor $Mn_{Ga}$-$Mn_{Ga}$ coupling. The general prediction of this more advanced formulation of the valence band model is that $T_C$ increases monotonically with $x_{eff}$ and $p$. Since in both theories the two key parameters are $p$ and $x_{eff}$ their experimental verification is hindered by the fact that accurate determination of the total hole concentration, especially in insulating samples, is quite difficult if one uses electrical measurements, since this contains little if any contribution from localized holes; and the anomalous Hall effect in ferromagnetic (Ga,Mn)As adds an additional level of complication. Furthermore, the value of $x_{eff}$ is usually determined from magnetization data, and this determination is based on an assumed value of magnetic moment per $x_{eff}$, which is not fully known.

In this paper we take a direct approach for determining both the effective Mn concentration $x_{eff}$ and the hole concentration $p$ from one experiment, namely, we map out the location of Mn in the lattice by simultaneously using channeling Rutherford backscattering (c-RBS) and channeling particle-induced X-ray emission (PIXE), which automatically yield the concentration of $Mn_{Ga}$ ($x_{sub}$) and $Mn_I$ ($x_I$), allowing us to directly evaluate $x_{eff}$ and $p$. These measurements were performed on a series of 10 samples, both as-grown and annealed, with a total Mn concentration $x_{tot}$ ranging between 0.03 and 0.068, i.e., on a wide range of samples having different degrees of compensation and transport behaviors that vary from insulating to metallic. We also performed transport, magnetization and magnetic circular dichroism (MCD) measurements on the same samples. This is the first time that such a detailed and systematic study of the location of Mn ions in the lattice and its correlation with other experiments has been carried out. Taken together, our results challenge the valence band picture of ferromagnetism in (Ga,Mn)As. Instead, our data point to the existence of a Mn-derived impurity band (IB) even in samples with $x_{tot}$ ~ 6.8%; and, moreover, they show that it is *the location of the Fermi level $E_F$ within IB*, rather than the hole concentration itself, that determines $T_C$ in (Ga,Mn)As. Specifically, samples with $E_F$ located among localized states are insulating and show low $T_C$, while samples with the $E_F$ located among the more extended states in the IB show high $T_C$ and a metallic behavior. Our results also indicate that, contrary to common belief [27, 34-37], $Mn_I$ are in fact necessary for achieving a high value of $T_C$ as predicted by Erwin and Petukhov[16]. Understanding their role is therefore crucial for guiding the growth and annealing conditions of (Ga,Mn)As.

The results of the channeling experiments and the corresponding values of $T_C$ are presented in Table I. The hole concentration p for a given sample in Table I was calculated as $p = 4(x_{sub} - 2x_I)/a^3$, where *a* is the lattice constant of (Ga,Mn)As in that sample calculated according to Ref. [38]. While this is not a direct measurement of *p*, it provides an excellent estimate, as demonstrated by Wojtowicz *et al.*[35], who used electrochemical capacitance-voltage profiling in (Ga,Mn)As to measure *p* directly, and compared it with the values obtained from channeling experiments. We note parenthetically that, although arsenic antisites ($As_{Ga}$) also act as double donors in (Ga,Mn)As, their concentration is typically below ~$4\times10^{19}$ cm$^{-3}$, and their effect on *p* is therefore negligible in samples with $x_{tot} > 0.03$ (i.e., all samples in this study)[39]. The value $p/N_{Mn}^{eff}$ appearing in the table represents the ratio of the hole concentration to the effective Mn concentration, a quantity relevant to the theoretical picture in Ref. [33], which is also equal to the filling factor $f = (x_{sub} - 2x_I)/x_{eff}$ introduced later in this paper. Our data confirm the well-known fact[27,34] that a large fraction of Mn ions in the as-grown samples reside at interstitial sites. Specifically, the ratio $x_I/x_{tot}$ is 6% for sample B; about 10% for samples C, and D, and 25% for sample F. Sample F is quite striking in this respect: although it has the highest total Mn concentration, its $x_{eff}$ is one of the lowest, and its *p* is the lowest of all samples because of the large value of $x_I$. Our data also confirm the well-known fact[30,40] that the main effect of annealing is the removal of $Mn_I$ (presumably into random sites).

The first conclusion which can be drawn from Table I is that the data are in striking disagreement with the valence band model prediction that $T_C$ increases monotonically with $x_{eff}$ and $p$. For example, sample F* has the highest $T_C$ of all samples, even though it has one of the lowest values of $x_{eff}$ and of $p$. In addition, sample D* has a $T_C$ comparable to that of sample F, but the latter has the lowest of all $p$ values and one of the lowest $x_{eff}$. A detailed comparison between our data and the prediction of the valence band model[33], which further illustrates these disagreements, is shown in Figure 1. The theory shown by the green line was compactly presented in Ref. [33] as a plot of $T_C/x_{eff}$ vs $p/N_{Mn}^{eff}$, the variable defined above, which can be seen as a measure of compensation level. (Low values of $p/N_{Mn}^{eff}$ indicate samples with high compensation level, and vice versa.) The red squares in the figure represent data from Table I, and the blue squares are taken from Ref. [41] for additional comparison. For samples from Ref. [41] the values of $x_{eff}$ were found from channeling experiments as well, while $p$ was taken from electrochemical capacitance-voltage profiling, a method that has been shown to agree quite well with $x_{sub} - 2x_I$.[35] As can be seen from the figure, the experimental data show a highly non-monotonic behavior, in sharp contrast to the valence band picture developed in Ref. [33]. Specifically, samples with a high degree of compensation are in agreement with the predictions of Ref. [33]; however, the lower is the compensation level (i.e., the larger the $p/N_{Mn}^{eff}$ ratio), the more drastic is the disagreement between the valence band picture and our experiment. We emphasize that this disagreement has only come to light through the simultaneously obtained c-RBS and c-PIXE data, the strength of this approach being that both $p$ and $x_{eff}$ are determined by applying *the same* experimental method to each sample. It is especially important to note that, since the values of $p$ and $x_{eff}$ obtained by this method are directly related to each other, the uncertainty in the key parameter of the theory involved, i.e., $p/N_{Mn}^{eff}$, is very small.

Further information is obtained from transport measurements. Figure 2 shows the resistivity $\rho$ as a function of temperature $T$ for samples C, D and F, both as-grown and annealed. All samples show lower resistivity upon annealing, as expected, but interestingly the samples showing a clearly metallic behavior (F and F*), have their hole concentrations among the lowest. In contrast, samples C, C*, D and D* show insulating behavior (seen especially clearly for sample D), although all these samples have higher $p$ than F and F*.

We now present the results of magnetic circular dichroism (MCD) measurements on the same set of samples. Earlier MCD studies on $Ga_{1-x}Mn_xAs$ with low Mn concentration ($x \approx$ 0.015) revealed that the MCD signal originates from the spin-dependent difference in the density of states (DOS) in the valence band brought about by the presence of a Mn-derived spin-polarized impurity band, rather than from the $p$-$d$ enhanced Zeeman splitting of the valence band[42]. This is consistent with recent findings that the $p$-$d$ exchange splitting in the valence band is negligibly small[22,25]. The model developed earlier[42] is based on the fact that the creation of an impurity state near $Mn_{Ga}$ results in a depletion of the valence band by a state with the spin oriented parallel to the Mn spin. Thus, when the IB is spin-polarized, fewer spin states parallel to the Mn spin are left in the valence band, as sketched in Fig. 3.

Figure 4 shows MCD spectra taken on six of the samples, in order of increasing $x_{eff}$. For comparison we also include MCD data for $x_{eff} = 0.016$[42]. All spectra show the same general features: a very broad positive signal that rises sharply at the energy gap (1.5 eV, indicated by the vertical line). Since the spectra for higher $x_{eff}$ evolve smoothly, showing similar features as the spectra for $x_{eff} = 0.016$, we conclude that the model developed earlier[42] also applies to higher $x_{eff}$, i.e., that MCD arises from a spin imbalance in the valence band states, a behavior that can only be attributed to the existence of a spin-polarized IB. Furthermore, since the MCD signal rises sharply *at the band gap*, not only must there be a difference in the spin-up and spin-down DOS at the top of the valence band, but these top valence band states must be occupied. Thus the Fermi level must lie above the top of the valence band, i.e., in the impurity band.

As was also shown in Ref. [42], in the case of very low Mn concentrations the largest contribution to impurity states comes from heavy holes (hh). However, as the number of states in the IB increases with increasing $x_{eff}$, the light hole states (lh) also begin to be pulled into the IB states. Figure 4 shows that as $x_{eff}$ increases, a second peak begins to develop at higher energy with systematically increasing intensity. We therefore ascribe the two peaks observed in MCD to peak contributions from the hh band (lower energy peak) and from lh band (higher energy peak).

While the existence of an IB for $x_{tot} \sim 0.068$ is sometimes questioned, we argue that it is actually $x_{eff}$ that controls the number of impurity states. The experimental evidence for this comes from Fig. 4, which reveals that samples with similar $x_{eff}$ display similar MCD spectra. Especially striking is the resemblance between the spectra for samples F and B, even though they have very different values of $x_{tot}$, $p$, and $T_C$ (see Table I). The only property which these samples have in common is their small $x_{eff}$. This behavior, together with the evolution of the lh peak in the MCD spectra with increasing $x_{eff}$, provides a clue that it is $x_{eff}$ rather than $x_{tot}$ or even $x_{sub}$ that controls the shape of MCD spectra.

Additional confirmation that $x_{eff}$ controls the number of impurity states is presented in Fig. 5, where we compare the MCD spectra for samples F and F*. As shown in Table I, both samples have very similar values of $x_{sub}$. However, $x_{eff}$ increases from 2.5% in F to 3.5% in F*. Since the MCD signal arises from the spin-dependent difference in the DOS in the valence band, and since the number of spin-up states missing from the valence band must equal the number of states of the same spin in IB, the integrated MCD signal over the energy range above the band gap must reflect the number of states in IB. The increase of the MCD signal in the annealed sample F* in Fig. 5 thus indicates that the number of states in IB is determined by $x_{eff}$. Other samples also show some increase in the integrated MCD upon annealing, but because annealing in those samples leads only to small increases of $x_{eff}$, the effect is not as dramatic as in F and F*.

We now discuss these findings in the context of the unsolved issue concerning the states occupied by the holes mediating the Mn-Mn ferromagnetic coupling. As already mentioned, our results disagree with the models based on the valence band picture. In addition, the MCD data for nominal Mn concentrations of up to ~6.8% are consistent with the model predicting that MCD arises from the spin imbalance in VB originating from the presence of a spin-polarized impurity band, and that $E_F$ lies above the top of the valence band in all our samples. Since the data also show that the number of states in the IB is determined by $x_{eff}$, the hole fraction $p/N_{Mn}^{eff}$ listed in Table I is recognized as the filling factor $f$ of IB, i.e, $f = (x_{sub} - 2x_I)/x_{eff}$, and thus gives the location of the Fermi level within the IB. For example, $f = 0.75$ indicates that 25% of the IB states are occupied by electrons, while $f = 0.9$ indicates that only 10% of the IB states are occupied by electrons, putting the Fermi level near the bottom of the IB.

Taking all our data together, we argue that the behavior of the Curie temperature of (Ga,Mn)As can be understood only by assuming that its ferromagnetism is mediated by holes residing in the impurity band, and that it is the location of the Fermi level within IB that determines $T_C$. We can now re-examine Figure 1 and see how $T_C/x_{eff}$ changes as the Fermi level moves from the top of the impurity band ($f \approx 0.2$) to the bottom ($f \approx 0.9$). As the figure shows, samples with $f \approx 0.7$ have the highest $T_C/x_{eff}$ ratio, and that ratio drops for lower as well as for higher values of $f$. This "dome-like" behavior of $T_C/x_{eff}$ vs $f$ agrees conceptually with the prediction that $T_C$ is expected to be highest at $f \approx 0.5$ in the IB, because ferromagnetic alignment is promoted by hole hopping which is only possible between singly occupied and empty Mn impurity states[16]. This result is also in agreement with the simulations of Ref. [43], which find extended states and the highest DOS close to the center of the IB, meaning that the most metallic samples with the highest $T_C$ have their Fermi level in this region. In contrast, for small or large $f$, i.e., when the Fermi level is near the top or the bottom of the IB, the DOS is lower and the states become localized, suggesting insulating samples with lower $T_C$. As an illustration, samples A and F* have their Fermi levels close to the center of IB; thus they have the highest $T_C/x_{eff}$ ratio, and are also most metallic. However, more detailed theoretical modeling is needed before quantitative comparisons can be undertaken.

Comparing samples with similar filling factors of $f \sim 0.9$ also provides a valuable insight into the effect of annealing. As shown in Fig. 2, the resistivities in annealed samples, especially C* and D*, decrease by about two orders of magnitude after annealing. Since $f$ also increases slightly, thus moving $E_F$ toward more localized states, we are forced to conclude that annealing lowers the disorder and improves the uniformity of the sample. As a consequence, the energy range where the states are extended is expanded, resulting in an increased $T_C$. This agrees with neutron reflection measurements[44] and ferromagnetic resonance experiments[45] which also show that one of the major effects of annealing is to improve the homogeneity of (Ga,Mn)As samples.

In summary, our data are consistent with the existence of an impurity band up to rather high nominal Mn doping levels, and indicate that the location of the Fermi level within the IB plays a crucial role in determining $T_C$ through determining the degree of localization of the IB holes.

Specifically, we show that having the Fermi level near the middle of the impurity band, where the states are most extended, is at least as important for raising $T_C$ as increases in $x_{eff}$. Understanding of the role of $E_F$ thus opens the pathways for new strategies for achieving higher values of $T_C$. For example, appropriate control of the concentration of $Mn_I$, co-doping with donor ions[41], or modulation doping can be used to engineer the location of $E_F$ within IB to best advantage. Similarly, the findings of this paper have far reaching consequences for optimizing the ferromagnetic coupling (and therefore $T_C$) within the whole family of III-Mn-V ferromagnetic semiconductors through tuning the binding energy of the Mn acceptors via, e.g., alloying different III-V host materials. By varying the binding energy of Mn acceptors one can vary the location and the width of the impurity band, and therefore improve the Curie temperature through improving the mobility of the IB holes.

## Methods

**Synthesis of (Ga,Mn)As samples**: (Ga,Mn)As samples with a thickness of ~100 nm were grown by low temperature molecular beam epitaxy (LT-MBE) on (001) GaAs substrates on which a 0.2 μm $Ga_{0.70}Al_{0.30}As$ buffer was first deposited. The annealed samples were treated at 280°C for one hour in flowing $N_2$ gas. Superconducting quantum interference device (SQUID) magnetometry was used to measure the magnetization of the samples as a function of temperature in order to determine $T_C$. These measurements were complemented by electrical transport characterization as a function of temperature.

**Magneto-optical spectroscopy:** The MCD measures the difference in absorption of right- and left- circularly-polarized light, and is given by the expression

$$MCD = \frac{T^+ - T^-}{T^+ + T^-} \sim \frac{(\alpha^- - \alpha^+)d}{2},$$

where $T^+$ and $T^-$ are transmission intensities for $\sigma^+$ and $\sigma^-$ circular polarizations, and $\alpha\pm$ are the corresponding absorption coefficients. Since the MCD measurements were performed in a transmission mode, the GaAs substrate had to be removed through polishing and subsequent etching, the $Ga_{0.70}Al_{0.30}As$ buffer serving as an excellent etch stop. The measurements of MCD were carried out using polarization modulation via a photo-elastic modulator. The samples were placed in an optical cryostat equipped with a 6 Tesla superconducting magnet, and the magnetic field was applied parallel to the direction of light propagation, i.e., normal to the (Ga,Mn)As layers. The transmitted light was detected by a photodiode using a lock-in amplifier.

**Channeling experiment:** The total Mn concentration and the locations of Mn sites in the $Ga_{1-x}Mn_xAs$ lattice were studied by simultaneous channeling Rutherford backscattering spectrometry (*c*-RBS) and channeling particle induced X-ray emission (*c*-PIXE) measurements using a 1.95 MeV $^4$He$^{++}$ beam. Backscattered He ions and characteristic X-rays excited by the He ions were detected by a Si surface barrier detector located at a backscattering angle of 165° and a Si(Li) detector located at 30° with respect to the incident ion beam. The specific locations of Mn atoms in the lattice were determined by directly comparing the angular scans about the <110> and <111> axial channels of the Mn K$_\alpha$ X-ray signals (PIXE) with those of the RBS signals from the GaAs host lattice. The accuracy of determining the quantities $x_{sub}$, $x_I$ and $x_{tot}$ relevant to the present paper is estimated as ±10%.

# REFERENCES

[1] Dietl, T. A ten-year perspective on dilute magnetic semiconductors and oxides. Nature Materials **9**, 965-974 (2010).

[2] Dietl, T., Ohno, H., Matsukura, F., Cibert, J. & Ferrand, D. Zener model description of ferromagnetism in zinc-blende magnetic semiconductors. Science **287**, 1019-1022 (2000).

[3] Jungwirth, T. et al. Character of states near the Fermi level in (Ga,Mn)As: impurity to valence band crossover. Phys. Rev. B **76**, 125206 (2007).

[4] Sawicki, M. Magnetic properties of (Ga,Mn)As. J. Magn. Magn. Mater. **300**, 1-6 (2006).

[5] Neumaier, D. et al. All-Electrical Measurements of the Density of States in (Ga,Mn)As. Phys. Rev. Lett. **103**, 087203 (2009).

[6] Sawicki, M. et al. Experimental probing of the interplay between ferromagnetism and localization in (Ga,Mn)As. Nature Phys. **6**, 22-25 (2010).

7       Richardella, A. et.al. Visualizing critical correlations near the metal-insulator transition in Ga$_{1-x}$Mn$_x$As. Science **327**, 665-669 (2010).

8       Boukari, H. et al. Light and electric field control of ferromagnetism in magnetic quantum structures. Phys. Rev. Lett. **88**, 207204 (2002).

9       Jungwirth, T., König, J., Sinova, J., Kučera, J. & MacDonald, A. H. Curie temperature trends in (III,Mn)V ferromagnetic semiconductors. Phys. Rev. B **66**, 012402 (2002).

10      Nishitani, Y. et al. Curie temperature versus hole concentration in field-effect structures of Ga$_{1-x}$Mn$_x$As. Phys. Rev. B **81**, 045208 (2010).

11      Wang, K. Y. et al. Influence of the Mn interstitial on the magnetic and transport properties of (Ga,Mn)As. J. App. Phys. **95**, 6512-6514 (2004).

12      Ku, K. C. et al. Highly enhanced Curie temperature in low-temperature annealed [Ga,Mn]As epilayers. Appl. Phys. Lett. **82**, 2302-2304 (2003).

13      Sato, K., Dederichs, P. H. & Katayama-Yoshida, H. Curie temperatures of III-V diluted magnetic semiconductors calculated from first principles. Europhys. Lett. **61**, 403–408 (2003).

14      Berciu, M. & Bhatt, R. N. Effects of disorder on ferromagnetism in diluted magnetic semiconductors. Phys. Rev. Lett. **87**, 107203 (2001).

15      Mahadevan, P. & Zunger, A. Trends in ferromagnetism, hole localization, and acceptor level depth for Mn substitution in GaN, GaP, GaAs, GaSb. Appl. Phys. Lett. **85**, 2860–2862 (2004).

16      Erwin, S. C. & Petukhov, A. G. Self-compensation in Manganese-doped ferromagnetic semiconductors. Phys. Rev. Lett. **89**, 227201 (2002).

17    Alberi, K. et al. Formation of Mn-derived impurity band in III-Mn-V alloys by valence band anticrossing. Phys. Rev. B **78**, 075201 (2008).

18    Mayer, M. A. et al. Electronic structure of $Ga_{1-x}Mn_xAs$ analyzed according to hole-concentration-dependent measurements. Phys. Rev. B **81**, 045205 (2010).

19    Burch, K., Awschalom, D. & Basov, D. Optical properties of III-Mn-V ferromagnetic semiconductors. J. Magn. Magn. Mater. **320**, 3207–3228 (2008).

20    Burch, K. S. et al. Impurity Band Conduction in a High Temperature Ferromagnetic Semiconductor. Phys. Rev. Lett. **97**, 087208 (2006).

21    Rokhinson, L. P. et al. Weal localization in $Ga_{1-x}Mn_xAs$: Evidence of impurity band transport. Phys. Rev. B **76**, 161201 (R) (2007).

22    Ohya, S., Muneta, I., Hai, P. N. &Tanaka, M. Valence-Band Structure of the Ferromagnetic Semiconductor (Ga,Mn)As Studied by Spin-Dependent Resonant Tunneling Spectroscopy. Phys. Rev. Lett. **104**, 167204 (2010).

23    Sheu, B. L. et al. Onset of Ferromagnetism in Low-Doped $Ga_{1-x}Mn_xAs$. Phys. Rev. Lett. **99**, 227205 (2007).

24    Tang, J.-M. & Flatte, M. E. Magnetic Circular Dichroism from the Impurity Band in III-V Diluted Magnetic Semiconductors. Phys. Rev. Lett. **101**, 157203 (2008).

25    Ohya, S., Takata, K. & Tanaka, M. Nearly non-magnetic valence band of the Ferromagnetic semiconductor (Ga,Mn)As. Nature Physics **7**, 342-347 (2011).

26    B. C. Chapler, B.C. et al. Infrared probe of the insulator-to-metal transition in $Ga_{1-x}Mn_xAs$ and $Ga_{1-x}Be_xAs$. Phys. Rev. B **84**, 081203 (R) (2011).

27    Yu, K. M. et al. Effect of the location of Mn sites in ferromagnetic $Ga_{1-x}Mn_xAs$ on its Curie temperature. Phys. Rev. B **65**, 201303(R) (2002).

28    Blinowski, J. & Kacman, P. Spin interactions of interstitial Mn ions in ferromagnetic (Ga,Mn)As. Phys. Rev. B **67**, 121204 (2003).

29    Mašek, C.J., & Máca, F. Interstitial Mn in (Ga,Mn)As: Binding energy and exchange coupling. Phys. Rev. B **69**, 165212 (2004)

30    Edmonds, K. W. et al. Mn Interstitial Diffusion in (Ga,Mn)As. Phys. Rev. Lett. **92**, 037201 (2004).

31    Bouzerar, G., Ziman, T., & Kudrnovský, J. Compensation, interstitial defects, and ferromagnetism in diluted ferromagnetic semiconductors. Phys. Rev. B **72,** 125207 (2005).

32    Takeda, Y. et al. Nature of Magnetic Coupling between Mn Ions in As-Grown $Ga_{1-x}Mn_xAs$ Studied by X-Ray Magnetic Circular Dichroism. Phys. Rev. Lett. **100**, 247202 (2008).

33    Jungwirth, T. et al. Prospects for high temperature ferromagnetism in (Ga.Mn)As semiconductors. Phys. Rev. B **72**, 165204 (2005).

34    Yu, K. M. et al. Curie temperature limit in ferromagnetic $Ga_{1-x}Mn_xAs$. Phys. Rev. B **68**, 041308(R) (2003).

35    Wojtowicz, T., Furdyna, J. K., Liu, X., Yu, K. M., Walukiewicz, W. Electronic effects determining the formation of ferromagnetic $III_{1-x}Mn_xV$ alloys during epitaxial growth. Physica E **25**, 171-180 (2004).

36    MacDonald, A. H., Schiffer, P. & Samarth, N. Ferromagnetic semiconductors: moving beyond (Ga,Mn)As. Nature Materials **4**, 195–202 (2005).

37      Yu, K. M. et al. Fermi Level Effects on Mn Incorporation in III-Mn-V Ferromagnetic Semiconductors. *Spintronics*, *Semiconductors and Semimetals*, Vol. 82, edited by Tomasz Dietl, David D. Awchalom, Maria Kamińska, Hideo Ohno (Elsevier, San Diego, 2008).

38      Sadowski, J. et al. Structural and magnetic properties of (Ga,Mn)As layers with high Mn-content grown by migration-enhanced epitaxy on GaAs(100) substrates. Appl. Phys. Lett. **78**, 3271-3273 (2001).

39      Wolos, A. et al. Properties of arsenic antisite defects in $Ga_{1-x}Mn_xAs$. J. Appl. Phys. **96**, 530-533 (2004).

40      Wang, K. Y. et al. Influence of the Mn interstitial on the magnetic and transport properties of (Ga,Mn)As. J. Appl. Phys. **95**, 6512-6514 (2004).

41      Cho, Y. J., Yu, K. M., Liu, X., Walukiewicz, W. & Furdyna, J. K. Effects of donor doping on $Ga_{1−x}Mn_xAs$. Appl. Phys. Lett. **93**, 262505 (2008).

42      Berciu, M. et al. Origin of magnetic circular dichroism in (Ga,Mn)As: giant Zeeman splitting vs. spin dependent density of states. Phys. Rev. Lett. **102**, 247202 (2009).

43      Moca, C. P., Zarand, G. & Berciu, M. Theory of optical conductivity for dilute $Ga_{1-x}Mn_xAs$. Phys. Rev. B **80**, 165202 (2009).

44      Kirby, B. J. et al. Annealing-Dependent Magnetic Depth Profile in $Ga_{1-x}Mn_xAs$. Phys. Rev. B **69**, 081307(R) (2004).

45      Liu, X., Sasaki,Y. & Furdyna, J. K. Ferromagnetic resonance in $Ga_{1-x}Mn_xAs$. Phys. Rev. B **67**, 205204 (2003).

**Acknowledgments**

K.T. thanks Dr. Y.-Y. Zhou for her help with the MCD set up and sample preparation. This work was supported by the National Science Foundation Grant DMR 10-05851; by NSERC and CIFAR and by the Director, Office of Science, Office of Basic Energy Sciences, Materials Sciences and Engineering Division, of the U.S. Department of Energy under Contract No. DE – AC02-05CH11231.

**Author contributions**

M.D and M.B conceived the project and wrote the manuscript. K.T carried out the MCD, transport and magnetization experiments with guidance from X.L, M.D and J.K.F. X.L fabricated the samples and contributed to the manuscript. K.M.Y and W.W are responsible for the channeling experiments. The project was supervised by M.D and J.K.F. All authors have reviewed, discussed and approved the results and conclusions of this article.

**Additional information**

The authors declare no competing financial interests. Correspondence and requests for materials should be addressed to M. Dobrowolska.

**FIGURE LEGENDS**

**Figure 1. Comparison between experimental data and theoretical calculations based on the valence band model of Jungwirth et al.[33]** The calculations according to Ref. [33] (shown by green curve) were carried out within TBA/CPA model based on the valence band picture, and include such effects as hole-hole exchange interaction, discreteness of random $Mn_{Ga}$ position in the lattice, and other refinements. The calculations and the data shown in the figure are plotted as $T_C/x_{eff}$ vs $p/N_{Mn}^{eff}$ (where $p/N_{Mn}^{eff}$ is the ratio of the hole concentration to the concentration of effective Mn moments $N_{Mn}^{eff}$, with $N_{Mn}^{eff} = 4x_{eff}/a^3$). The theoretical calculations of Ref. [33] predict that for high compensation the Curie temperature falls with decreasing hole concentration. In the small compensation regime, however, these predictions show that $T_C$ is almost independent of the hole density. The experimental data shown in the figure are taken from Table I (red squares) and from Ref. [41] (blue squares). For samples in the low compensation range the experimental points are seen to depart drastically from the theoretical predictions of the valence band model. The error bars in the figure are calculated using standard analysis of error propagation, i.e., $\delta f^2 = \sum_i^n (\partial f/\partial x_i)^2 (\delta x_i)^2$, where the error bars are given by $\delta f$, and $\delta x_i$ are the standard deviations of $Mn_I$, $Mn_{Ga}$ and $T_C$.

**Figure 2. Effect of annealing on transport properties of** (Ga,Mn)As **samples.** The resistivity ρ was measured as a function of temperature for as grown (black curves) and annealed (red curves) (Ga,Mn)As films: a) Samples C and C*, b) Samples D and D*, and c) Samples F and F*.

**Figure 3. Schematic representation of the origin of magnetic circular dichroism (MCD) in (Ga,Mn)As.** Band structure of (Ga,Mn)As is shown for two spin orientations (not to scale; for more details see discussion in Supplementary Information). The creation of an impurity state near a substitutional Mn ion results in the depletion of the valence band (VB) by a state with spin oriented parallel to the Mn spin. When the impurity band is fully spin-polarized, there are fewer spin-down states than spin-up states left in VB. Consequently the absorption due to valence-band-to-conduction-band transitions for the $\sigma^+$ circular polarization (marked $\alpha^+$) is weaker than that for the opposite circular polarization ($\alpha^-$), as indicated by the thickness of the long arrows. This spin-dependent difference in the density of states (DOS) in VB leads then to a positive MCD signal that begins at the energy gap. Since the DOS vanishes at the band-edges, this spin-dependent difference in the DOS will increase with energy, reaching a maximum where the spin-down DOS in VB is depleted the most by contributions to the IB, and again decreasing at higher energies. The light hole band (not included in the figure for clarity) makes a similar contribution to MCD at higher energies.

**Figure 4. Magnetic circular dichroism (MCD) spectra.** The spectra shown are taken at temperature T = 10 K and magnetic field B = 5 T. Since the thickness of the layers is the same for all samples ($d \sim 100$ nm), the MCD signal $(T^+ - T^-)/(T^+ + T^-)$ expressed in % facilitates good quantitative comparison between the samples. The vertical line shown at 1.5 eV indicates the band gap of GaAs. The top spectrum, obtained for a sample with effective Mn concentration $x_{eff}$ = 1.6%, was published earlier[42], and is included here to show that the spectra obtained for samples with higher Mn concentrations evolve smoothly from low $x_{eff}$ onward and show the same general features: a very broad positive signal that rises sharply at the band gap. The peak which develops on the higher energy side and increases in intensity as $x_{eff}$ increases reflects the increasing contribution to the impurity band from the light hole band.

**Figure 5. Comparison between magnetic circular dichroism (MCD) spectra for samples F and F*.** The spectra were taken at $T = 10K$ and magnetic field $B = 5T$. The integrated intensity of the MCD signal for the annealed sample F* is significantly higher (~40%) than for the corresponding as-grown sample F. A below-gap negative contribution to the MCD signal is seen for sample F, as discussed in Supplementary Information.

**Table I**

| Sample | $x_{tot}$ | $x_{random}$ | $x_{sub}$ | $x_I$ | $x_{eff}$ | $T_C$ | $p$ ($10^{20}$ cm$^3$) | $p/N_{Mn}^{eff}$ or $f$ |
|---|---|---|---|---|---|---|---|---|
| A (as grown) | 3.0% | 0.1% | 2.4% | 0.5% | 1.9% | 45 K | 3.1 | 0.74 |
| B (as grown) | 3.5% | 0.1% | 3.2% | 0.2% | 3.0% | 27 K | 6.2 | 0.93 |
| C (as grown) | 4.8% | 0.3% | 4.0% | 0.5% | 3.5% | 34 K | 6.6 | 0.86 |
| C* (annealed) | 5.0% | 0.4% | 4.3% | 0.3% | 4.0% | 54 K | 8.2 | 0.93 |
| D (as grown) | 5.8% | 0.3% | 5.0% | 0.5% | 4.5% | 26 K | 8.8 | 0.90 |
| D* (annealed) | 5.9% | 0.4% | 5.3% | 0.2% | 5.1% | 42 K | 10.8 | 0.96 |
| E (as grown) | 6.0% | 0.3% | 5.0% | 0.7% | 4.3% | 30 K | 8.0 | 0.84 |
| E* (annealed) | 6.1% | 0.5 % | 5.2% | 0.4% | 4.8% | 43 K | 9.7 | 0.92 |
| F (as grown) | 6.4% | 0.5% | 4.2% | 1.7% | 2.5% | 40 K | 1.8 | 0.32 |
| F* (annealed) | 6.8% | 1.3% | 4.5% | 1.0% | 3.5% | 90 K | 5.5 | 0.71 |

**Table I. Local structure of (Ga,Mn)As as determined by channeling experiments.** The location of Mn sites in the lattice was determined by simultaneous channeling-particle-induced x-ray emission (*c*-PIXE) and Rutherford backscattering spectrometry (*c*-RBS). Here $x_{tot}$ indicates the total Mn concentration; $x_{random}$ the concentration of Mn atoms residing at random locations, e.g., MnAs inclusions; $x_{sub}$ the concentration of Mn ions located at Ga lattice sites (Mn$_{Ga}$); and $x_I$

is the concentration of Mn ions located at interstitial sites ($Mn_I$). The quantity $x_{eff} = x_{sub} - x_I$ is the concentration of Mn ions which contributes to ferromagnetic order. The hole concentration $p$ for a given sample was found as $p = 4(x_{sub} - 2x_I)/a^3$, where $a$ is the lattice constant of (Ga,Mn)As of that sample calculated according to Ref. [38]. The Curie temperature $T_C$ is determined from magnetization measurements vs temperature. Finally, $p/N_{Mn}^{eff}$ is the number of holes per effective Mn moment $N_{Mn}^{eff} = 4(x_{eff})/a^3$. It is easily seen that $p/N_{Mn}^{eff}$ is equal to the filling factor of the impurity band $f = (x_{sub} - 2x_I)/x_{eff}$ used later in discussing the location of the Fermi level. Even though there is a general tendency for $x_I/x$ to increase in samples with higher Mn concentration, in practice its value is a very sensitive function of growth conditions, i.e., the probability for $Mn_I$ to form depends on both the Mn flux and the growth temperature.

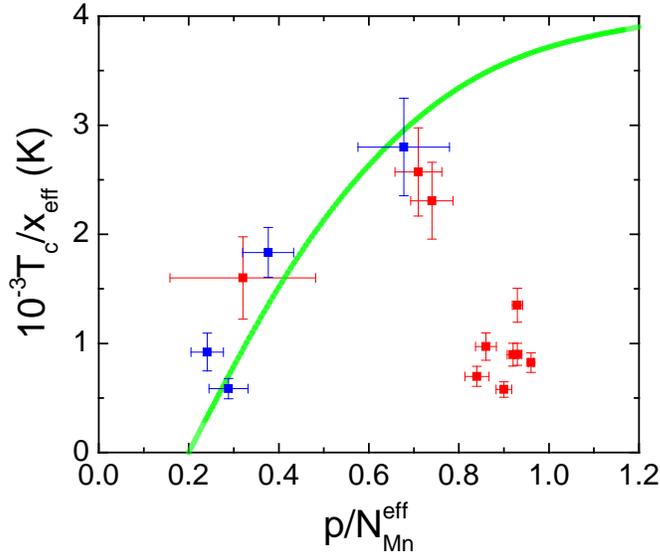

Figure 1

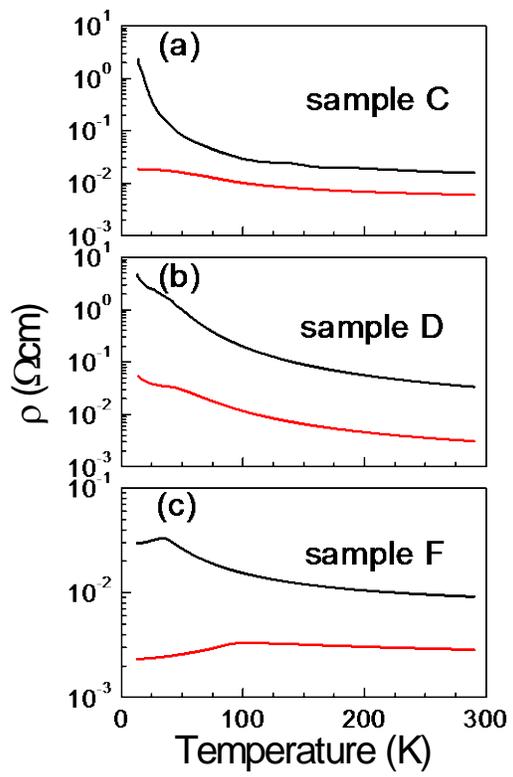

Figure 2

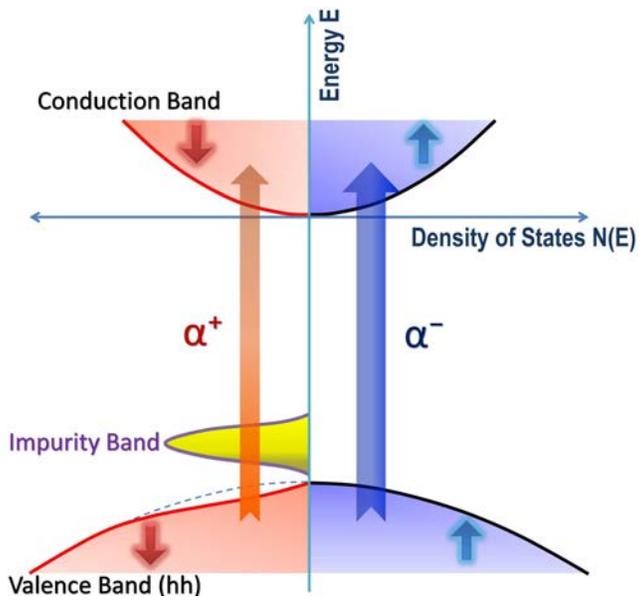

Figure 3

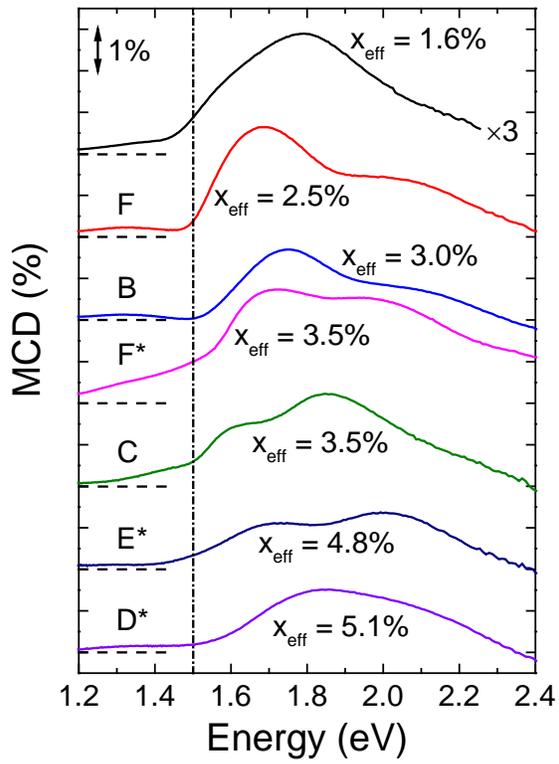

Figure 4

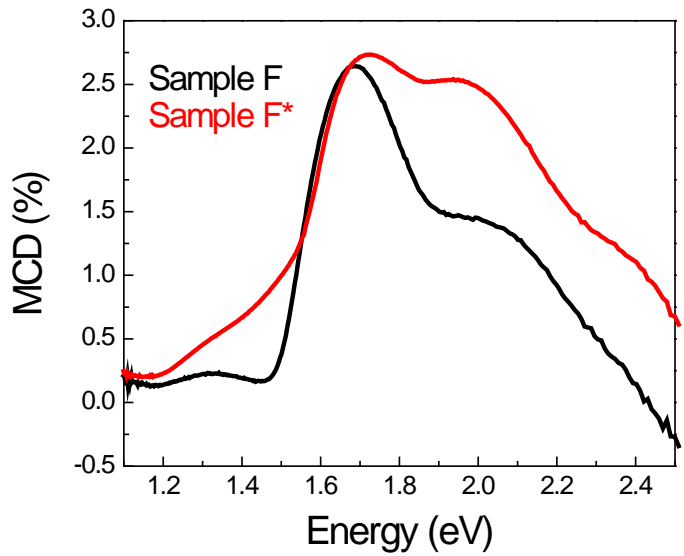

Figure 5

**Supplementary information for the article "Controlling Curie temperature in (Ga,Mn)As through location of the Fermi level within the impurity band", by**
M. Dobrowolska, K. Tivakornsasithorn, X. Liu, J. K. Furdyna, M. Berciu, K. M. Yu and W. Walukiewicz

**S.1. Spin dependent density of states, magnetic circular dichroism, and depletion of the valence band**

One should note that Figure 3 in the main text, which is a schematic representation of the mechanism of magnetic circular dichroism (MCD) in (Ga,Mn)As, is drawn with exaggeration for pedagogical reasons. If we had plotted this figure to scale, it would be very difficult to see any effects arising from the spin-dependent depletion of the valence band that causes MCD. Specifically, as can be seen in Figure 5 in the main text, the value of MCD at its maximum does not exceed 3%. Since the MCD signal is proportional to the density of states (DOS) pulled from the valence band into the impurity band[1], we can conclude that not more than 3% of the states are missing from the valence band. This estimate also agrees with our assertion that the number of states in the impurity band (IB), and therefore the depletion of one of the spin projections in the VB, is of the order of $x_{eff}$. Thus, one should not expect any significant changes in the shape of the valence band.

It is worth mentioning that the data in Fig. 2D of Richardella et al.[2] supports such depletion of the VB density of states with increasing $x_{eff}$. Indeed, the differential conductance *dI/dV* plotted in that figure is believed to be proportional to the DOS. Their curves clearly show that the DOS well inside the VB (indicated by the y-axis intercept of the plot) decreases with increasing $x_{tot}$. Since the entire VB is not shown in the above figure, we cannot estimate what percentage of the DOS is moved into the IB according to their measurements; but the trend is certainly consistent with our findings.

Results reported in the work of Ohya et al.[3] are also consistent with our findings. If the Mn ions bind the impurity states in their immediate vicinity, it follows that the (somewhat fewer) states left in the VB will have little overlap with the Mn sites, because the VB states are orthogonal to the impurity states. This also follows from the completeness of the basis of eigenstates. Since the impurity states make a large contribution to the density of states near the Mn ion, the VB states must contribute little in those regions. Because of this very small overlap with the Mn ion, the VB states are quite insensitive to the magnetic properties of Mn, thus resulting in very small *p-d* exchange. In fact, apart from some depletion in their number and some redistribution of their wavefunctions so as to avoid the regions occupied by impurity states, the VB states are similar to those in a non-magnetic material. This is one of the main observations of the paper by Ohya et al.[3], indicating that the VB is very insensitive to any magnetic properties of the material, including $T_C$. This also explains why the energies of the discrete VB levels trapped in the very thin (Ga,Mn)As films used by those authors are well described by the *k-dot-p* parameters of the non-magnetic GaAs host.

## S.2. Negative contribution to MCD signal below the energy gap

One of the important aspects of the MCD signal in (Ga,Mn)As is that it is comprised of both positive and negative contributions. Earlier studies of the MCD in (Ga,Mn)As with low Mn concentration showed a negative contribution to the MCD signal in some of the samples. In Ref.

[1]it was argued that this negative below-gap contribution arises due to compensation. As shown in Figure 4 of that reference, samples A, B and D studied in that paper have the same Mn concentration $x = 0.014$, but were grown under different conditions, resulting in different levels of compensation. In that series, sample D was most highly compensated, and the negative contribution to MCD in that sample was therefore very pronounced; while Sample A had the lowest degree of compensation, and the negative contribution to MCD signal was completely absent for that sample. In this context, it is interesting to compare the MCD spectra of samples A and B in Ref. [1] below the energy gap. Sample B has somewhat higher degree of compensation than sample A, and one can see in Fig. 4 of that paper that, even though the MCD signal in sample B never reaches the "negative territory", the contribution of compensation to the spectrum is clearly seen in the form of a downward dip in the spectrum of Sample B.

In the present paper a negative contribution to MCD signal is not expected because most of the samples used have a low compensation level, as indicated by their filling factors listed in Table I. The only sample with relatively high compensation level is sample F, for which $f \approx 0.3$, indicating that 70% of the states in the impurity band are occupied by electrons. Figure 5 of the present manuscript compares MCD spectra on samples F (as-grown) and F*(annealed). As one can see, and in agreement with Fig. 4 of Ref. 1 discussed in the preceding paragraph, the negative contribution below the gap is clearly evident (again in the form of a downward dip), even though the signal never actually drops below zero. It is difficult to say whether there is any negative contribution to MCD below the gap in sample F* ($f = 0.7$) because of the presence of a positive below-gap background. However, based on our understanding, such negative contribution (if it exists) would in this case be even weaker than in sample F.

References:

[1] Berciu, M. et al. Origin of magnetic circular dichroism in (Ga,Mn)As: giant Zeeman splitting vs. spin dependent density of states. Phys. Rev. Lett. **102**, 247202 (2009).

[2] Richardella, A. et.al. Visualizing critical correlations near the metal-insulator transition in $Ga_{1-x}Mn_xAs$. Science **327**, 665-669 (2010).

[3] Ohya, S., Takata, K. & Tanaka, M. Nearly non-magnetic valence band of the Ferromagnetic semiconductor (Ga,Mn)As. Nature Physics **7**, 342-347 (2011).